%% file: main.tex
\documentclass{article}
\usepackage{microtype}
\usepackage{graphicx}
\usepackage{subfigure}
\usepackage{booktabs}
\usepackage{amsmath, amssymb}
\usepackage{bm}
\usepackage{xcolor}
\usepackage{multirow}
\usepackage[T1]{fontenc}
\usepackage{hyperref}

\usepackage[accepted]{icml2020}
\usepackage{pdfpages}
\icmltitlerunning{When deep denoising meets iterative phase retrieval}

\begin{document}
\twocolumn[
\icmltitle{When deep denoising meets iterative phase retrieval}
\begin{icmlauthorlist}
\icmlauthor{Yaotian Wang}{elepu}
\icmlauthor{Xiaohang Sun}{elepu}
\icmlauthor{Jason W. Fleischer}{elepu}
\end{icmlauthorlist}
\icmlaffiliation{elepu}{Department of Electrical Engineering, Princeton University, Princeton, New Jersey, USA}
\icmlcorrespondingauthor{Jason W. Fleischer}{jsonf@princeton.edu}
\icmlkeywords{}
\vskip 0.3in
]
\printAffiliationsAndNotice{}  
\begin{abstract}
Recovering a signal from its Fourier intensity underlies many important applications, including lensless imaging and imaging through scattering media. Conventional algorithms for retrieving the phase suffer when noise is present but display global convergence when given clean data. Neural networks have been used to improve algorithm robustness, but efforts to date are sensitive to initial conditions and give inconsistent performance. Here, we combine iterative methods from phase retrieval with image statistics from deep denoisers, via regularization-by-denoising. The resulting methods inherit the advantages of each approach and outperform other noise-robust phase retrieval algorithms. Our work paves the way for hybrid imaging methods that integrate machine-learned constraints in conventional algorithms.
\end{abstract}
\input{introduction.tex}

\input{background.tex}

\input{relatedworks.tex}

\input{methods.tex}
\input{experiment.tex}

\input{conclusion.tex}

\bibliographystyle{icml2020}
\end{document}


\maketitle

In this supplementary material, we provide proofs on the proximal operators used in our algorithms and show how ADMM \cite{boyd2011distributed} with indicator functions coincides with Hybrid-Input-Output (HIO) \cite{fienup1982phase} and Hybrid-Projection-Reflection (HPR) \cite{bauschke2003hybrid}.
\section{Proximal operators}
We consider two proximal operators for Fourier phase retrieval: the squared error of Fourier amplitudes and regularization by denoising (RED) coupled with additional object-space constraints.
\begin{enumerate}
    \item $R(x) = \bar{I}_\mathcal{C}(x) + \frac{\lambda}{2}\langle x, x-D(x)\rangle$\\
    
Let $D$ be the denoiser used in RED and $\mathcal{C}$ be the set of signals satisfying the additional constraints provided, where we assume that the denoiser $D$ is (locally) homogeneous with symmetric Jacobian \cite{romano2017little} and $\mathcal{C}$ is a convex set. For any $\tau>0$, if $v^+ = \text{prox}_{\tau R}(v)$, then the first-order optimality condition gives
\begin{equation}\label{REDnCcondition}
\begin{split}
\;&v^+ = \underset{x\in \mathbb{R}^n}{\text{armin}}\;\tau R(x)+ \frac{1}{2}\rVert v - x\rVert^2\\
\overset{}{\Rightarrow}\;& \tau( \partial \bar{I}_\mathcal{C}(v^+) + \lambda(v^+ - D(v^+)) + v^+ -v = 0\\
\Leftrightarrow\;&v^+ =  \left(I + \frac{\tau}{1+\lambda\tau}\partial \bar{I}_\mathcal{C}\right)^{-1}\left(\frac{v+\lambda \tau D(v^+)}{1+\lambda\tau}\right)\\
\Leftrightarrow\;&v^+ =  \Pi_\mathcal{C}\left(\frac{v+\lambda \tau D(v^+)}{1+\lambda\tau}\right)
\end{split}
\end{equation}
where $\partial \bar{I}_\mathcal{C}$ is the subgradient of the indicator function and the last equality follows by noting that the resolvent of $\partial \bar{I}_\mathcal{C}$ is the projection $\Pi_\mathcal{C}$ onto $\mathcal{C}$ \cite{ryu2016primer}.
\\
\\
\item $f(z) = \frac{1}{2}\rVert y - |Fz|\rVert^2$\\

Let $F$ be the (normalized) discrete Fourier transform and $y$ be the measured Fourier amplitude, which is non-negative. For simplicity, we consider 1D signals only (the conclusion holds for any dimension). Using the overhead symbol $\hat{\cdot}$ to denote the signal after Fourier transform, Parseval's theorem gives
\begin{equation}\label{def:prox_f}
\begin{split}
    &x^+=\text{prox}_{\tau f}(x)  = \text{argmin}_z\; \frac{\tau}{2} \rVert y - |Fz|\rVert_2^2 + \frac{1}{2}\rVert x -z\rVert^2 \\
    \Leftrightarrow\; &\widehat{x^+}=\text{argmin}_{\hat{z}}\; \frac{\tau}{2} \rVert y - |\hat{z}|\rVert_2^2 + \frac{1}{2}\rVert \hat{x} -\hat{z}\rVert^2 \\
    \; &\quad\;= \text{argmin}_{\hat{z}}\;\frac{1}{2} \sum_{k}\tau(|\hat{z}[k]| - y[k])^2 +  |\hat{z}[k]-\hat{x}[k]|^2
\end{split}
\end{equation}

It was noticed in \cite{wen2012alternating} that the solution is
\begin{equation}
\widehat{x^+}[k] = \frac{\tau}{\tau + 1}y[k]\frac{\hat{x}[k]}{|\hat{x}[k]|} + \frac{1}{\tau + 1}\hat{x}[k] \quad \forall k
\end{equation}
which follows from the first-order optimality condition. Here, we provide an alternative proof that this solution is the global minimum.

We start by using the triangle inequality $|\hat{z}[k] - \hat{x}[k]|^2 \geq \left( |\hat{z}[k]| - |\hat{x}[k]|\right)^2$ to give the lower bound 

\begin{equation}
    \min_{\hat{z}}\;\sum_{k}\tau(|\hat{z}[k]| - y[k])^2 +  |\hat{z}[k]-\hat{x}[k]|^2 \geq \min_{\hat{z}}\;\sum_{k}\tau(|\hat{z}[k]| - y[k])^2 +  \left( |\hat{z}[k]| - |\hat{x}[k]|\right)^2
\end{equation}
Equality between the right- and left-hand sides is achieved when
\begin{equation}
    \Re(\overline{\hat{z}[k]}\hat{x}[k]) = |\hat{z}[k]\hat{x}[k]|\quad \forall k
\end{equation}
i.e., when the complex phase $\angle \hat{z}[k] = \angle \hat{x}[k]$ ($\angle \hat{z}[k]$ can be arbitrary if $\hat{x}[k]=0$). As the right-hand side is convex on $|\hat{z}[k]|$, the minimum is achieved when
\begin{equation}
    |\hat{z}[k]| = \frac{\tau y[k] + |\hat{x}[k]|}{\tau + 1}\quad \forall k
\end{equation}
as $y[k], |\hat{x}[k]|\geq 0$. Therefore, if $x^+$ minimizes (\ref{def:prox_f}), then for all $k$,
\begin{equation}
\begin{split}
       \widehat{x^+}[k] &= \frac{\tau y[k] + |\hat{x}[k]|}{\tau + 1}\exp(i\angle \hat{x}[k]) \\
       &= \frac{\tau}{\tau + 1}y[k]\exp(i\angle \hat{x}[k]) + \frac{1}{\tau + 1}|\hat{x}[k]|\exp(i\angle \hat{x}[k])\\
       &= \frac{\tau}{\tau + 1}y[k]\frac{\hat{x}[k]}{|\hat{x}[k]|} + \frac{1}{\tau + 1}\hat{x}[k]
\end{split}
\end{equation}
Performing an inverse Fourier transform gives (26) in the main text:
\begin{equation}
    {x^+} = \frac{\tau}{1+\tau}\Pi_\mathcal{M}(x) + \frac{1}{\tau + 1}x
\end{equation}
\\
\end{enumerate}

\section{Equivalence between ADMM and HIO\textbackslash HPR}
Let $x_0$ be the ground truth and $S$ and $\tilde{S}$ be the support for $x_0$ and the extended support for padded $\tilde{x}_0 = P_{mn}x_0$, respectively. 

If there is additional information about the signal support, e.g. an estimation $\gamma$ such that $S\subseteq \gamma$, then the relation $\tilde{S}\subseteq \tilde{\gamma}$ holds for the extended support as well. For example, if we use the same vectorization as in the main text, such that
\begin{equation}
    \tilde{x} = P_{mn}x = \begin{bmatrix}
    x\\
    0_{m-n}
    \end{bmatrix}
\end{equation}
then we will have $S = \tilde{S}$ and $\gamma = \tilde{\gamma}$. Define subset $\mathcal{S}$ for the signals satisfying the given support constraint,
\begin{equation}
    \mathcal{S}:=\{x\in\mathbb{C}^n \;|\;x_i=0\;\forall i\notin \gamma\}
\end{equation}
The projection onto $\mathcal{S}$ is
\begin{equation}
    \Pi_\mathcal{S}(x)_i = \begin{cases}
    x_i &\text{if }i\in \gamma\\
    0 &\text{otherwise}
    \end{cases}
\end{equation}
and similarly for $\tilde{\mathcal{S}}:=\{x\in\mathbb{C}^m \;|\;x_i=0\;\forall i\notin \tilde{\gamma}\}$ on the extended support.

According to \cite{bauschke2002phase}, HIO with $\beta =1$ can be written as
\begin{equation}\label{HIO_general}
    \tilde{x}^{k+1} = \Pi_{\tilde{\mathcal{S}}}(2\Pi_\mathcal{M}(\tilde{x}^{k}) -  \tilde{x}^{k}) -  \Pi_{{\mathcal{M}}}(\tilde{x}^{k}) + \tilde{x}^{k}
\end{equation}

We now relate this to the optimization of FPR with the support constraint
\begin{equation}
\begin{split}
        \underset{x\in\mathbb{C}^n,z\in\mathbb{C}^m}{\text{minimize}}\;&\bar{I}_{\mathcal{M}}(z) + \bar{I}_\mathcal{S}(x)\\
        \text{subject to }&z = O_{mn}x
\end{split}
\end{equation}
With $\tilde{x} = O_{mn}x$, this can be rewritten as
\begin{equation}\label{fpr_optim}
\begin{split}
        \underset{\tilde{x},z\in\mathbb{C}^m}{\text{minimize}}\;&\bar{I}_{\mathcal{M}}(z) + \bar{I}_\mathcal{\tilde{S}}(\tilde{x})\\
        \text{subject to }&z = \tilde{x}
\end{split}
\end{equation}
for which ADMM gives the update rule as
\begin{equation}\label{ADMM_update}
    \begin{split}
        \tilde{x}^{k+1} &= \Pi_{\tilde{\mathcal{S}}}(z^k + u^k)\\
        z^{k+1} &= \Pi_\mathcal{M}(\tilde{x}^{k+1} -u^k) \\
        u^{k+1} &= u^k +  z^{k+1} - \tilde{x}^{k+1}
    \end{split}
\end{equation}
As in \cite{wen2012alternating}, the updates for ${m}^{k+1} = \tilde{x}^{k+1} - u^k$ are given by
\begin{equation}\label{ADMM_transformed}
    \begin{split}
        {m}^{k+2} &= \tilde{x}^{k+2} - u^{k+1}
        \\
    &= \Pi_{\tilde{\mathcal{S}}}(2\Pi_\mathcal{M}(m^{k+1})-m^{k+1}) - \Pi_\mathcal{M}(m^{k+1}) + m^{k+1}
    \end{split}
\end{equation}
which coincides with (\ref{HIO_general}). 

Next, we denote $\mathcal{S}_+$ as the set containing signals which not only satisfy the support constraint but also have non-negative elements in the real part:
\begin{equation}
    \mathcal{S}_+ := \{x\in\mathbb{C}^n\;|\;x_i=0\;\forall i \notin \gamma\text{ and }\Re(x_i)\geq 0\; \forall i\}
\end{equation}
The projection onto $\mathcal{S}_+$ is
\begin{equation}
    \Pi_{\mathcal{S}_+}(x) = \Pi_{Re_+}\left(\Pi_{\mathcal{S}}(x)\right)
\end{equation}
with $\Pi_{Re_+}$ being the element-wise projection 
\begin{equation}
    \Pi_{Re_+}(x)_i = \begin{cases}
    x_i &\text{if }\Re(x_i)\geq 0\\
    i\Im(x_i) &\text{otherwise}
    \end{cases}
\end{equation}
Changing $\mathcal{S}$ to $\mathcal{S}_+$ in (\ref{fpr_optim}) and repeating (\ref{ADMM_update}) to (\ref{ADMM_transformed}) gives the recursion for $m^{k+1}$ as
\begin{equation}
    {m}^{k+2} = \Pi_{\tilde{\mathcal{S}}_+}(2\Pi_\mathcal{M}(m^{k+1})-m^{k+1}) - \Pi_\mathcal{M}(m^{k+1}) + m^{k+1}
\end{equation}
which coincides with HPR with $\beta = 1$ \cite{bauschke2003hybrid}.
\\

\bibliographystyle{icml2020}

%% file: introduction.tex
\section{Introduction}
\label{intro}
In computational imaging, numerical algorithms are used to estimate a signal ${{x}\in} \mathbb{R}^n$ or $\mathbb{C}^n$ from raw data ${y}$ (generally obtained from a physical system). One of the most common computational imaging schemes is Phase Retrieval (PR), in which ${x}$ is retrieved through the phaseless measurements of the output of a linear system
\begin{equation}\label{general-measurement}
    y^2 = M(x) + w = |Ax|^2 + w 
\end{equation}
where $A$ is a known linear transform and $w$ is the noise in the measurements. In the past decade, the general phase retrieval (PR) problem has attracted much attention from the optimization and statistics community \cite{candes2015phasemc,candes2015phasewf, wang2017solving, chen2017solving}. Despite a solid theoretical foundation, general algorithms have overly restrictive requirements (e.g. the statistics of measurement bases) that have limited their popularity. More progress has been made for Fourier phase retrieval (FPR), in which $A$ is the result of transformed or far-field measurements. This is also the most common type experimentally, with applications ranging from astronomy \cite{fienup1987phase} to diffraction \cite{miao1999extending, chapman2010coherent} and speckle-correlation \cite{bertolotti2012non, katz2014non} imaging.

The most broadly used algorithms for FPR are iterative methods, pioneered by Gerchberg-Saxton \cite{gerchberg1972practical} and later developed by Fienup \cite{fienup1982phase}. Though they lack theoretical proof of convergence, empirical use of Fienup algorithms and their variants \cite{bauschke2003hybrid, elser2003phase, luke2004relaxed, martin2012noise, rodriguez2013oversampling} has shown the avoidance of local minima and convergence to global solutions from random initialization. Together with the simplicity of their implementation, iterative phase retrieval methods have become the workhorse of FPR \cite{miao2005quantitative, bertolotti2012non, katz2014non}.  

It has been shown that applying a natural image prior to FPR can increase robustness to noise and improve reconstruction quality \cite{venkatakrishnan2013plug, heide2016proximal, metzler2018prdeep, Isil:19}. However, such methods either have unsatisfying robustness when noise levels are high or are sensitive to initialization (thus relying on other algorithms to supply initial points). Both cases return us to the problem of poor reliability when the signal-to-noise ratio in measurements is low. 

Our major contribution here is to combine the benefits of iterative FPR with natural image priors via Regularization-by-Denoising (RED) \cite{romano2017little}. The methods we propose deliver greater robustness to noise than other noise-robust FPR algorithms while relaxing the initialization requirements. The application of image priors also alleviates the stagnant mode issues in iterative phase retrieval \cite{fienup1986phase}, leading to accelerated convergence. Machine learning thus resolves long-lasting issues that have hindered traditional methods. In turn, traditional algorithms can lift the burden on deep learning by focusing it on a subset of the whole, end-to-end problem.

%% file: background.tex
\section{Background}
We focus on two-dimension signals and assume the measurement transform $A$ in (\ref{general-measurement}) to be the (normalized) Fourier transform
\begin{equation}
    \hat{x}[k_1,k_2] = \frac{1}{\sqrt{n}}\sum_{n_1, n_2=0}^{\sqrt{n}}x[n_1,n_2]e^{-2\pi i\frac{n_1k_1+n_2k_2}{\sqrt{n}}}
\end{equation}
Below, we discuss the uniqueness of Fourier phase retrieval, common algorithms used, and their relation to more general optimization problems.

\subsection{Uniqueness in FPR}
If there is not enough sampling, the Fourier intensity may be insufficient to trace back to the input signal. For all $d$-dimensional signals with $d\geq 2$, except a set of measure 0 \cite{hayes1982reducible}, it has been shown that if the Fourier intensity is oversampled by a factor greater than 2 in each dimension, then a signal is determined uniquely by its Fourier intensity up to the trivial ambiguities of translation, conjugate inversion and global phase \cite{hayes1982reconstruction}. Fortunately, in practice these ambiguities are often acceptable, since the geometrical transform and global phase keep the characteristics of the object intact.

Oversampling in the Fourier domain is related to the so-called support constraint for FPR, which is a more often used terminology in iterative phase retrieval. For example,   suppose the Fourier spectrum of $x\in\mathbb{R}^{\sqrt{n}\times\sqrt{n}}$ is oversampled twice uniformly at $k_i = \{0, 1/2, 1, \cdots, \sqrt{n}-1/2\} = \frac{1}{2}\{0,1,\cdots, 2\sqrt{n}-1\}=\frac{1}{2}\tilde{k}_i$ for $i=1,2$, which is denoted as $\hat{x}^{(2)}$. By defining $\tilde{x}\in\mathbb{C}^{\sqrt{m}\times \sqrt{m}}$ with $m = 4n$ such that $\tilde{x}[n_1, n_2]=\sqrt{\frac{m}{n}}x[n_1, n_2]$ if $n_i\in \mathbb{N}<\sqrt{n}$ and $\tilde{x}[n_1, n_2]=0$ otherwise, we have
 \begin{equation}
 \begin{split}
 \hat{x}^{(2)}[k_1,k_2] &= \frac{1}{\sqrt{n}} \sum_{n_1,n_2=0}^{\sqrt{n}}{x}[n_1,n_2]e^{-i2\pi\frac{n_1k_1+n_2k_2}{\sqrt{n}}}\\
 &= \frac{1}{\sqrt{n}}\sum_{n_1,n_2=0}^{\sqrt{m}}\sqrt{\frac{n}{m}}\tilde{x}[n_1,n_2]e^{-i2\pi\frac{n_1\tilde{k}_1+n_2\tilde{k}_2}{\sqrt{m}}}\\
 &=\hat{\tilde{x}}[\tilde{k}_1,\tilde{k}_2]
\end{split}
\end{equation}
where $\hat{\tilde{x}} = F\tilde{x}$, with $F$ being the 2D DFT transform on vectorized signal in $\mathbb{C}^m$ and $F^* = F^{-1}$ being the inverse transform. Therefore, there exists a supported signal $\tilde{x}$ by zero-padding $P_{mn}$ and scaling $x$ by a factor of $\sqrt{m/n}$, such that its Fourier transform is the same as (uniform) oversampling in the Fourier space of $x$. If the vectorization order gives
\begin{equation}
    \tilde{x}^T = \sqrt{\frac{m}{n}}\begin{bmatrix}
    x^T & 0^T_{m-n}
    \end{bmatrix}
\end{equation}
then 
\begin{equation}
    \hat{x}^{(2)} = F\tilde{x} = FO_{mn}x
\end{equation}
where $O_{mn}\in \mathbb{R}^{m\times n}$ is given by
\begin{equation}
    O_{mn} = \sqrt{\frac{m}{n}}\begin{bmatrix}
    I_{n} \\ 0
    \end{bmatrix} = \sqrt{\frac{m}{n}}P_{mn}
\end{equation}
Stated another way, oversampling FPR is equivalent to finding a supported signal $\tilde{x}$ from its DFT intensity, with the support constraint sometimes including the support of $x$ itself. To distinguish them, we denote the support for $x\in \mathbb{C}^n$ as $S=\{i\;|\;x_i\neq 0\}$ and the extended support for padded $\tilde{x}$ as $\tilde{S}=\{j\;|\;\tilde{x}_j\neq 0\}$
\subsection{ADMM}
The Alternating Direction Method of Multipliers (ADMM) \cite{boyd2011distributed} is a popular algorithm for solving the linear constrained optimization problem
\begin{equation}
\begin{split}
    \underset{x_1,\cdots,x_N}{\text{minimize}} \quad&\ell(x_1,\cdots,x_N) = \sum_i^Nf_i(x_i)\\
    \text{subject to} \quad& \sum_i^N A_ix_i = b
\end{split}
\end{equation}
For each iteration, ADMM updates each $x_i$ and dual variable $u$ independently as
\begin{equation}\label{admm-update}
    \begin{split}
        x_i^{k+1} &= \underset{x_i}{\text{argmin}}f_i(x_i) + \frac{\rho^k}{2}\left\rVert \sum_{j\neq i} A_jx^{k}_j+A_ix_i-b+u^k\right\rVert^2\\
        u^{k+1} &= u^{k} + \sum_{i=1}^N A_ix^{k+1}_i -b
    \end{split}
\end{equation}
with penalty parameter $\rho^k$ being constant or adaptive through iterations. 

One often needs to evaluate the minimization problem of a form
\begin{equation}
    z^+ = \text{argmin}_{{v}} \; f({v}) + \frac{1}{2}\rVert {v-z}\rVert^2_2
\end{equation}
which is defined as the proximal operator for $f$ and $z$, i.e. $\text{prox}_{f}(z) = z^+$ \cite{parikh2014proximal}. The efficiency of ADMM generally depends on the complexity of evaluating the proximal operator for each $f_i$, while in return the functions can be non-differentiable. We show below that this latter property can be quite beneficial.
\subsection{Hybrid-Input-Output method}
As possibly the most used iterative method in FPR, the Hybrid-Input-Output (HIO) \cite{fienup1982phase} is well-known for its ability to converge to global minima from random initialization. HIO iterates on the padded and scaled signal $\tilde{x}$ with following step rules:
\begin{equation}
\begin{split}
    \tilde{z}^{k+1} &= {F}^{-1}\left( {y}\odot\frac{{F\tilde{x}^{k}}}{|{F\tilde{x}^{k}}|}\right) \\
   \forall i,\; \tilde{x}^{k+1}_i &= \begin{cases}
    \tilde{z}^{k+1}_i &\text{if }i\in \tilde{S}\\
    \tilde{x}^{k}_i - \beta \tilde{z}^{k+1}_i &\text{otherwise}
    \end{cases}
\end{split}
\end{equation}
where $\odot$ is the element-wise product. 

It was shown in \cite{bauschke2002phase} that HIO with $\beta = 1$ coincides with Douglas-Rachford splitting (DRS) \cite{douglas1956numerical, lions1979splitting, eckstein1992douglas}. Since DRS is equivalent to ADMM updates on the feasibility problem with indicator functions \cite{boyd2011distributed}, one can find that HIO ($\beta=1$) is equivalent to ADMM on the following minimization problem:
\begin{equation}\label{hio_admmform}
\begin{split}
    \underset{x\in \mathbb{C}^n, z\in \mathbb{C}^m}{\text{minimize}}\; &\bar{I}_\mathcal{M}(z) +\bar{I}_\mathcal{C}(x)\\
    \text{subject to}\;&z = O_{mn}x
\end{split}
\end{equation}
where the indicator function for a subset $\mathcal{S}$ is defined as \cite{boyd2004convex}
\begin{equation}
    \bar{I}_\mathcal{S}({x}) =
    \begin{cases}
    0 &\text{if }{x}\in \mathcal{S}\\
    +\infty &\text{otherwise}
    \end{cases}
\end{equation}
and the set $\mathcal{M}$ is defined as the set of signals consistent with the measurement
\begin{equation}\label{def:setM}
    \mathcal{M}:= \{{x}\in\mathbb{C}^m \;|\;|Fx| = y\}
\end{equation}
Here, $\mathcal{C}$ is the set of signals satisfying an additional constraint, such as inset support $S$ and nonnegativity (which result in the Hybrid-Projection-Reflection algorithm \cite{bauschke2003hybrid}). More details of this mapping are given in the supplementary material.

The indicator function (12) has the proximal operator as the projection to the corresponding set
\begin{equation}
    \Pi_\mathcal{S}({x}) := \underset{{{z}\in \mathcal{S}}}{\text{argmin}} \;\rVert {z-x}\rVert = \text{prox}_{\bar{I}_\mathcal{S}}({x}) 
\end{equation}
In particular, the projection onto $\mathcal{M}$ can be written as
\begin{equation}\label{proj_M}
    {F}^{-1}\left( {y}\odot\frac{{Fv}}{|{Fv}|}\right) \in \Pi_\mathcal{M}({v}) = \text{prox}_{\bar{I}_\mathcal{M}}(v)
\end{equation}

%% file: relatedworks.tex
\section{Related Works}
In this section, we introduce the efforts to date for solving the PR problem in the presence of noise.
\subsection{Iterative phase retrieval}
Iterative phase retrieval methods commonly solve a feasibility problem, looking for a signal whose oversampled Fourier intensity is $y^2$ and simultaneously is consistent with the other constraint $\mathcal{C}$. The problem occurs when noise levels increase in the measurement, resulting in oscillations and ambiguous solutions. To alleviate the degradation from corrupted data, efforts have been made to limit the effect of noise on iterative methods \cite{luke2004relaxed, martin2012noise, rodriguez2013oversampling}. However, without further priors on the object space (e.g. image statistics), the denoising effect of these methods is often insufficient.

\subsection{Deep learning in PR}
Deep neural networks (DNN) are well-known for their capability to approximate complicated functions (given enough training data). In image processing, they have achieved significant improvements over traditional methods in areas such as denoising \cite{zhang2017beyond, zhang2018ffdnet}, deblurring \cite{nimisha2017blur}, and superresolution \cite{dong2014learning,lim2017enhanced}. For solving PR, forward deep networks have shown some success in end-to-end predictions \cite{sinha2017lensless, rivenson2018phase}, while network-assisted algorithms also have helped in support estimation \cite{kim2019fourier}, low-light \cite{goy2018low} and compressive \cite{hand2018phase} situations.

However, using a forward neural network to approximate the inverse mapping is problematic for oversampling FPR. Such methodology relies on the assumption that forward mapping is one-to-one and well-posed; this is not the case here with even precise knowledge of the signal support, due to the existence of trivial ambiguities. Instead, the optimization method commonly adopted for solving FPR (e.g., in \cite{heide2016proximal, metzler2018prdeep}) minimizes the loss function
\begin{equation}\label{Tikhonov-Wiener}
    \ell({x}) := f({x};{y}) + \alpha R({x})
\end{equation}
where $f$ is the data fidelity term and $R$ is a regularizer involving prior belief, e.g. natural image statistics. This method is effectively a maximum a posteriori \cite{venkatakrishnan2013plug}.

\subsection{Prior by denoisers}
Using a denoiser as the prior $R$ in (\ref{Tikhonov-Wiener}) has been proposed to boost image inference in inverse problems. There have been two major strategies to utilize the denoiser: Plug-and-Play (PnP) regularization \cite{venkatakrishnan2013plug} and Regularization-by-Denoising (RED) \cite{romano2017little}. In PnP methods, the proximal operator for an implicit regularizer $R$ is approximated by an image denoiser. This approach provides promising results both empirically \cite{venkatakrishnan2013plug, heide2014flexisp, heide2016proximal, metzler2016denoising, meinhardt2017learning, zhang2017learning} and theoretically \cite{chan2016plug}. Meanwhile, RED is a framework that constructs explicit regularizers with denoisers $D$ as the inner product between a signal and the noise it contains,
\begin{equation}\label{red}
    R({x}) = \frac{\lambda}{2}\langle {x}, {x} - D({x})\rangle
\end{equation}
It has been shown in \cite{romano2017little} that if the denoiser $D$ has the properties of (local) homogeneity and Jacobian symmetry, then evaluation of the proximal operator in (\ref{red}) requires the solution of
\begin{equation}\label{red_condition}
    {z} - {x} + \lambda ({z} - D({z})) = 0
\end{equation}
Though these properties rarely hold for common denoisers, Equation (\ref{red_condition}) can still be adopted either as an approximation or if certain conditions hold \cite{reehorst2018regularization}. Recent applications of RED to PR have demonstrated a significant boost in noise robustness compared with bare iterative methods \cite{metzler2018prdeep, wu2019online}.

%% file: methods.tex
\begin{figure*}
    \centering
    \includegraphics[width=1\linewidth]{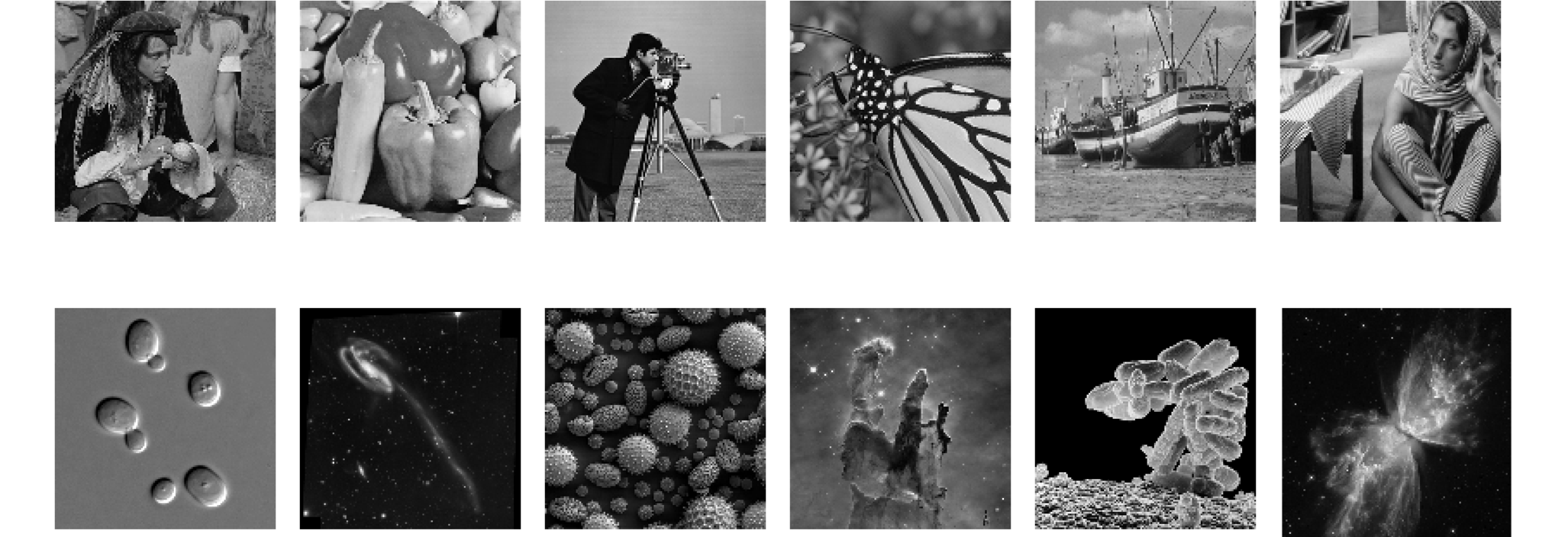}
    \caption{Test images used in the simulation. Top row: 6 commonly used ``natural'' test images \cite{zhang2017beyond}. Bottom row: 6 ``unnatural'' images \cite{metzler2018prdeep}. Images have been resized to $128\times 128$.}
    \label{fig:testimages}
\end{figure*}
\section{Methodology}

We aim to maintain the convergence benefits of HIO while alleviating the deleterious effects of noise. To this end, we adopt ADMM as a solver but modify the loss function used in HIO. More specifically, we eliminate the inconsistency from (\ref{hio_admmform}) by relaxation of the loss function and include natural image priors via RED, due to its explicit form and inherent flexibility.

For relaxation of the loss function, we consider two approaches: one in the Fourier constraint and one in the oversampling constraint. These result in two algorithms, RED-ITA-F and RED-ITA-S, respectively.

In general, we refer to our algorithms as RED-ITA, and Deep-ITA for the specific choice of deep denoisers, such as DnCNN \cite{zhang2017beyond}.

\subsection{RED-ITA-F}
We first consider substituting the indicator function on Fourier measurement to the data fidelity term. Following \cite{metzler2018prdeep}, we seek to solve
 \begin{equation}\label{loss-function-prdeep}
    \frac{1}{2} \rVert y - |FO_{mn}x|\rVert^2 + \frac{\lambda}{2}\langle x, x-D(x)\rangle
\end{equation}
 Similar to HIO, we transform (\ref{loss-function-prdeep}) into a linearly constrained form as
\begin{equation}\label{loss_ITA-F}
\begin{split}
    \underset{x\in\mathbb{R}^n,z\in\mathbb{C}^{m}}{\text{minimize}}\;&\frac{1}{2}\rVert y-|Fz|\rVert^2 + R(x)\\
    \text{subject to}\;& z = O_{mn}x
\end{split}
\end{equation}
where $R$ contains RED and an additional constraint $\bar{I}_{\mathcal{C}}(x)$:
\begin{equation}\label{regularizer}
    R(x) =  \bar{I}_{\mathcal{C}}(x)  +  \frac{\lambda}{2}\langle x, x- D(
    x)\rangle
\end{equation}
For $f({z}) =  \frac{1}{2}\rVert {y} - |{Fz}|\rVert^2$, the update rule of ADMM gives
\begin{equation}\label{update_RED-ITA-F}
\begin{split}
        x^{k+1} &= \underset{x\in \mathbb{R}^n}{\text{armin}}\;R(x)+ \frac{\rho}{2}\rVert z^k - O_{mn}x + u^k\rVert^2\\
        z^{k+1} &=\text{prox}_{\frac{1}{\rho}f}(O_{mn}x^{k+1} - u^k)\\
        u^{k+1} &= u^{k} + z^{k+1} - O_{mn}x^{k+1}
\end{split}
\end{equation}

It remains to evaluate each update step. We note that for any $x\in\mathbb{R}^n, v\in\mathbb{C}^m =\begin{bmatrix}
v_n^T & v_{m-n}^T,
\end{bmatrix}^T$ where $v_n = P^T_{mn}v$, 
\begin{equation*}
\begin{split}
&\rVert v - O_{mn}x \rVert^2 = \rVert \Re(v) - O_{mn}x\rVert^2 + \rVert \Im(v)\rVert^2\\
&=  \frac{m}{n}\left\rVert\sqrt{\frac{n}{m}}\Re(v_n)-x\right\rVert^2+ \rVert\Re(v_{m-n})\rVert^2 +\rVert \Im(v)\rVert^2
\end{split}
\end{equation*}
where $\Re(\cdot)$ and $\Im(\cdot)$ are the real and imaginary parts of a complex-valued signal. Therefore, in terms of $v = z^k + u^k$, the $x$-update step in (\ref{update_RED-ITA-F}) can be found as
\begin{equation}
\begin{split}
    x^{k+1} &= \underset{x\in \mathbb{R}^n}{\text{armin}}\;R(x)+ \frac{\rho}{2}\rVert v - O_{mn}x\rVert^2\\
    &= \underset{x\in \mathbb{R}^n}{\text{armin}}\;R(x)+ \frac{m\rho}{2n}\left\rVert \sqrt{\frac{n}{m}}\Re(v_n) - x\right\rVert^2\\
    &= \text{prox}_{\frac{n}{m\rho}R}(\sqrt{\frac{n}{m}}\Re(v_n))
\end{split}
\end{equation}
which reduces to an evaluation of the proximal operator for $R$. For any $\tau>0$, if $s^+ = \text{prox}_{\tau R}(s)$, we have
\begin{equation}\label{REDnCcondition}
s^+ =  \Pi_\mathcal{C}\left(\frac{s+\lambda \tau D(s^+)}{1+\lambda\tau}\right)
\end{equation} (a derivation is given in the supplementary material). Similar to RED in \cite{romano2017little}, the proximal operator in (\ref{REDnCcondition}) can be evaluated by the fixed-point approach, updating
\begin{equation}
    s^{(k+1)} =  \Pi_\mathcal{C}\left(\frac{s+\lambda \tau D(s^{k})}{1+\lambda\tau}\right)
\end{equation}
until convergence. In practice, the fixed point can be approximated by stopping after $p$ iterations with $s^{(0)} = s$, which is denoted as $\hat{\text{prox}}_{\tau R}(v) = s^{(p)}$ with $p$ being a hyperparameter. Empirically, we found that $p=1$ is efficient enough; therefore, $p$ is set to 1 in all of our experiments.

For the $z$-update step, the proximal operator for $f$ can be written as 
\begin{equation}
    \text{prox}_{\tau f}({s}) = \frac{1}{\tau + 1}{s} + \frac{\tau}{\tau + 1}\Pi_\mathcal{M}({s}) 
\end{equation}

This method for solving oversampling FPR is shown in Algorithm\ref{RED-ITA-F}. 
\begin{algorithm}[tb]
   \caption{RED-ITA-F}
   \label{RED-ITA-F}
\begin{algorithmic}
   \STATE {\bfseries Input:} Initialization $z^0, u^0\in\mathbb{C}^m$, $\rho,\lambda>0$, oversampling transform $O_{mn}$, Fourier measurement $y$
   \FOR{$k=0, 1, 2,\cdots$ }
   \STATE $v^k = z^k + u^k$
   \STATE $\tau = (m\rho)^{-1}n$
   \STATE $x^{k+1} = \hat{\text{prox}}_{\tau R}({\frac{n}{m}}\Re(O_{mn}^Tv^k))$
   \STATE $\tilde{x}^{k+1} = O_{mn}x^{k+1}$
   \STATE $z^{k+1} = \frac{\rho}{\rho+1}(\tilde{x}^{k+1} -u^k) + \frac{1}{\rho+1}\Pi_\mathcal{M}(\tilde{x}^{k+1} -u^k)$
   \STATE $u^{k+1} = u^{k} + z^{k+1} - \tilde{x}^{k+1}$
   \ENDFOR
\end{algorithmic}
\end{algorithm}
\subsection{RED-ITA-S}
The second approach is to relax the oversampling constraint, instead of the Fourier measurement. Rather than assuming there exists $x\in\mathbb{R}^n$ such that $O_{mn}x = z\in\mathcal{M}$, we acknowledge that the difference $\xi = z-O_{mn}x$ can be non-zero $\forall z\in\mathcal{M}$ and minimize the norm of it. That is, an alternative to (\ref{loss_ITA-F}) is
\begin{equation}\label{loss_ITA-S}
    \begin{split}
        \underset{z,\xi\in\mathbb{C}^m, x\in\mathbb{R}^n}{\text{minimize}}&\bar{I}_\mathcal{M}(z) + \frac{1}{2}\rVert \xi\rVert^2 + R(x)\\
        \text{subject to: }&z = O_{mn}x+\xi
    \end{split}
\end{equation}
Note that, given $x\in\mathbb{R}^n$, the loss in (\ref{loss_ITA-S}) is an upper bound for that in (\ref{loss_ITA-F}) since $\forall\;z\in\mathcal{M}$, Parseval's theorem gives
\begin{equation}
\begin{split}
    \rVert \xi\rVert^2 &= \rVert z - O_{mn}x \rVert^2 \\
    &= \rVert ye^{i\phi_{\hat z}} - FO_{mn}x\rVert^2\\
    &\geq \rVert y - |FO_{mn}x|\rVert^2
\end{split}
\end{equation}
where $\phi_{\hat{z}}$ is the Fourier phase of $z$.

A three-block ADMM is adopted to solve (\ref{loss_ITA-S}):
\begin{equation}\label{update_RED-ITA-S}
    \begin{split}
         x^{k+1} &= \underset{x\in \mathbb{R}^n}{\text{argmin}}\;R(x)+ \frac{\rho}{2}\rVert z^k - O_{mn}x -\xi^k+ u^k\rVert^2\\
        z^{k+1} &= \text{prox}_{\bar{I}_{\mathcal{M}}}\left(  O_{mn}x^{k+1} +\xi^k - u^k\right)\\
        \xi^{k+1} &=  \text{prox}_{\frac{1}{2\rho}\rVert \cdot\rVert^2}\left(z^{k+1} - O_{mn}x^{k+1} + u^k \right)\\
        u^{k+1} &= u^{k} + z^{k+1} - O_{mn}x^{k+1} - \xi^{k+1}
    \end{split}
\end{equation}
where
\begin{equation}
    \begin{split}
        \text{prox}_{\bar{I}_{\mathcal{M}}}(s) &= \Pi_\mathcal{M}(s)\\
        \text{prox}_{\frac{1}{2\tau}\rVert \cdot\rVert^2}(s) &= \frac{\tau s}{1+\tau}
    \end{split}
\end{equation}
This yields the RED-ITA-S shown in Algorithm \ref{RED-ITA-S}.

\begin{algorithm}[tb]
   \caption{RED-ITA-S}
   \label{RED-ITA-S}
\begin{algorithmic}
   \STATE {\bfseries Input:} Initialization $\xi^0, z^0, u^0\in\mathbb{C}^m$, $\rho,\lambda>0$, oversampling transform $O_{mn}$, Fourier measurement $y$
   \FOR{$k=0, 1, 2,\cdots$ }
   \STATE $v^k = z^k + u^k - \xi^k$
   \STATE $\tau = (m\rho)^{-1}n$
   \STATE $x^{k+1} = \hat{\text{prox}}_{\tau R}({\frac{n}{m}}\Re(O_{mn}^T v^k))$
   \STATE $\tilde{x}^{k+1} = O_{mn}x^{k+1}$
   \STATE $z^{k+1} \in \Pi_\mathcal{M}(\tilde{x}^{k+1} + \xi^k - u^k)$
   \STATE $\xi^{k+1} = \frac{\rho}{\rho + 1}\left(z^{k+1} - \tilde{x}^{k+1} + u^k \right)$
   \STATE $u^{k+1} = u^{k} + z^{k+1} - \tilde{x}^{k+1} - \xi^{k+1}$
   \ENDFOR
\end{algorithmic}
\end{algorithm}

\subsection{Connection between PR algorithms}
\cite{metzler2018prdeep} proposed solving (\ref{loss-function-prdeep}) with FASTA \cite{goldstein2014field}, a method known as prRED (the variant using deep denoisers like DnCNN is referred to prDeep). Since FASTA is a forward-backward splitting method, if the {{}stepsize} $\mu$ is fixed to be $n/m$ and $\lambda \rightarrow 0$, prDeep reduces to (sub-)gradient descent on the squared loss on Fourier amplitude, which coincides \cite{marchesini2007phase} with the Error Reduction algorithm \cite{fienup1982phase}.

RED-ITA-F reduces to HIO with $\beta=1$ when $\rho\rightarrow 0$ and $\lambda/\rho\rightarrow 0$. Similarly for DnCNN-ADMM if the denoising step is put first and $\mathcal{D}_\sigma$ is the identity transformation for $\sigma = 0$.

%% file: experiment.tex
\section{Experimental Results}
\begin{figure}[t]
    \centering
    \includegraphics[width=1\linewidth]{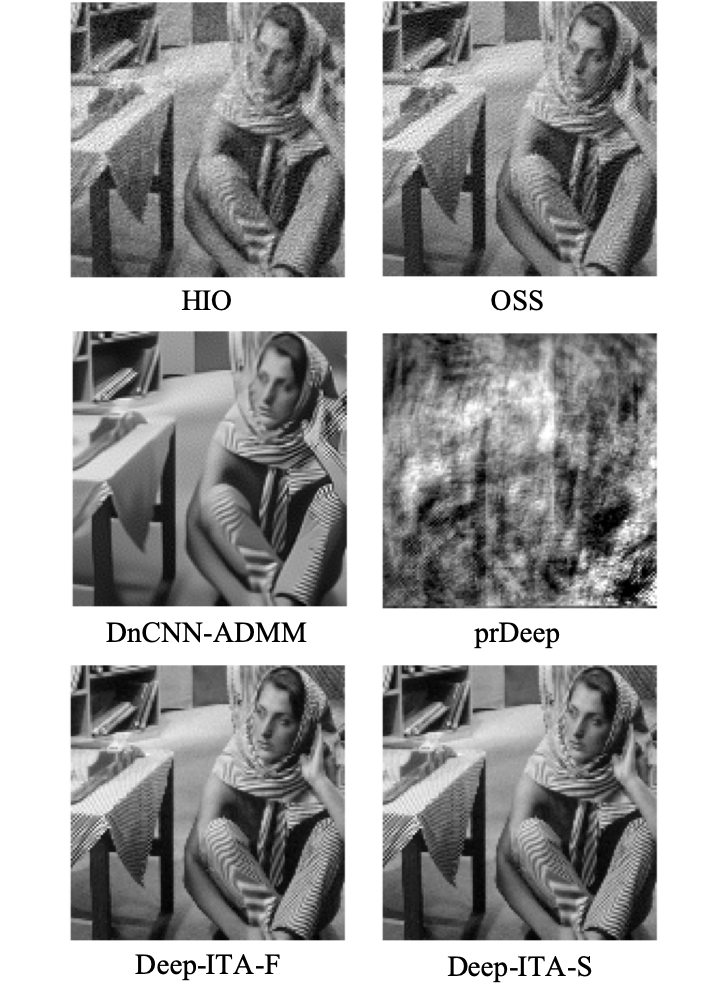}
    \caption{Reconstructions from random initialization with $\alpha = 4$. Both RED-ITA-F/S have the best reconstruction results.}
    \label{fig:rand_alpha4}
\end{figure}
\begin{figure}[t]
    \centering
    \includegraphics[width=1\linewidth]{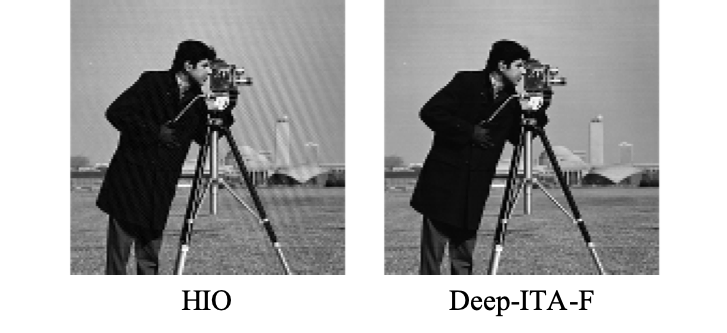}
    \caption{Reconstructions from random initialization with $\alpha = 0$. The stripes in the HIO reconstruction are artifacts from stagnation \cite{fienup1986phase}; they are resolved in our method.}
    \label{fig:rand_alpha0_stagnation}
\end{figure}
\begin{table*}[h!]
\caption{PSNRs and SSIMs of reconstructions initialized with random noise with varying noise level in the measurements. For $\alpha=0$, no noise is added to the Fourier intensity. For $\alpha = 4$, averaged MSNR$_1 = 32.09$dB, MSNR$_2 = 33.36$dB.}
\label{table_initializedrandom}
\vskip 0.15in
\begin{center}
\begin{small}
\begin{sc}
\begin{tabular}{l*{3}{r}*{3}{r}}
\toprule
 \multirow{2}{*}{$\alpha=0$}   & \multicolumn{3}{c}{Average PSNR}& \multicolumn{3}{c}{Average SSIM} \\
    & Natural &Unnatural& Overall & Natural&Unnatural&Overall\\
\hline
HIO & 48.88 & 56.01& 52.45 &0.94 & 0.88 &0.91 \\
OSS     & 24.27 & 44.31 & 34.29 &0.73 &0.82 &0.77\\
prDeep   & 13.70 &18.27 & 15.99 & 0.21 & 0.27 &0.24 \\
DnCNN-ADMM & 29.11 &27.94 & 28.52 & 0.87 & 0.74 &0.80 \\
DEEP-ITA-F   & \textbf{65.06} &57.88 & \textbf{61.47} & \textbf{1.00} & 0.99 &1.00 \\
DEEP-ITA-S & {64.94} & {\textbf{57.93}}& { {61.44}} & {{1.00}}&  {\textbf{0.99}} & \textbf{1.00}     \\
\bottomrule
\toprule
 \multirow{2}{*}{$\alpha=4$}   & \multicolumn{3}{c}{Average PSNR}& \multicolumn{3}{c}{Average SSIM} \\
    & Natural &Unnatural& Overall & Natural&Unnatural&Overall\\
\hline
HIO & 23.29 & 26.36& 24.83 &0.69 & 0.67 &0.68 \\
OSS     & 22.02 & 30.70 & 26.36 &0.64 &0.70 &0.67\\
prDeep   & 14.52 &19.46 & 16.99 & 0.23 & 0.36 &0.30 \\
DnCNN-ADMM & 28.46 &27.65 & 28.05 & 0.86 & 0.69 &0.77 \\
Deep-ITA-F   & 36.32 &34.39& 35.35 & 0.97 & {0.87} &{0.92} \\
Deep-ITA-S & \textbf{36.47} & \textbf{35.65}& \textbf {36.06}& \textbf{0.97}&  \textbf{0.91} & \textbf{0.94}      \\
\bottomrule
\end{tabular}
\end{sc}
\end{small}
\end{center}
\vskip -0.1in
\end{table*}
\begin{table*}[h!]
\caption{PSNRs and SSIMs of reconstructions initialized from HIO with varying noise level in the measurements. For $\alpha=8$, the averaged MSNR$_1 = 29.09$dB, MSNR$_2 = 27.54$dB. For $\alpha = 12$, averaged MSNR$_1 = 27.38$dB, MSNR$_2 = 24.49$dB. For $\alpha = 16$, averaged MSNR$_1 = 25.84$dB, MSNR$_2 = 22.52$dB.}
\label{table_initializedHIO}
\vskip 0.15in
\begin{center}
\begin{small}
\begin{sc}
\begin{tabular}{l*{3}{r}*{3}{r}}
\toprule
 \multirow{2}{*}{$\alpha=8$}   & \multicolumn{3}{c}{Average PSNR}& \multicolumn{3}{c}{Average SSIM} \\
    & Natural &Unnatural& Overall & Natural&Unnatural&Overall\\
\hline
HIO (init.)     & 20.78 & 23.03& 21.91 &0.56 & 0.53 &0.55 \\
OSS     & 22.02 & 27.58 & 24.80 &0.63 &0.65 &0.64\\
prDeep   & 28.50 &30.75& 29.62 & 0.87 & 0.79 &0.83 \\    
DnCNN-ADMM   & 26.95 &27.76& 27.35 & 0.81 & 0.68 &0.75 \\
Deep-ITA-F   & 32.90 &31.36& 32.13 & 0.94 & 0.83 &0.89 \\
Deep-ITA-S & \textbf{33.31} & \textbf{32.78}& \textbf {33.04}& \textbf{0.94}&  \textbf{0.86} & \textbf{0.90}      \\
\bottomrule
\toprule
 \multirow{2}{*}{$\alpha=12$}   & \multicolumn{3}{c}{Average PSNR}& \multicolumn{3}{c}{Average SSIM} \\
    & Natural &Unnatural& Overall & Natural&Unnatural&Overall\\
\hline
HIO (init.)     & 19.36 & 21.66& 20.51 & 0.47 & 0.45 &0.46 \\
OSS     & 20.78 & 25.07 & 22.93 &0.56 &0.56 &0.56\\
prDeep   & 28.24 &27.46& 27.85 & 0.85 & 0.74 &0.79 \\
DnCNN-ADMM   & 25.43 &25.89& 25.66 & 0.79 & 0.61 &0.70 \\
Deep-ITA-F   & 30.09 &29.11& 29.60 & 0.91 & 0.79 &0.85 \\
Deep-ITA-S & \textbf{31.95} & \textbf{30.38}& \textbf{31.17}& \textbf{0.93}&  \textbf{0.81} & \textbf{0.87}      \\
\bottomrule
\toprule
 \multirow{2}{*}{$\alpha=16$}   & \multicolumn{3}{c}{Average PSNR}& \multicolumn{3}{c}{Average SSIM} \\
    & Natural &Unnatural& Overall & Natural&Unnatural&Overall\\
\hline
HIO (init.)     & 17.59 & 20.50& 19.05 &0.36 & 0.38 &0.37 \\
OSS     & 19.65 & 23.37 & 21.51 &0.50 &0.51 &0.50\\
prDeep   & 26.44 &24.65 & 25.54 & 0.81 & 0.65 &0.73 \\
DnCNN-ADMM   & 22.87 &24.27 & 23.57 & 0.65 & 0.55 &0.60 \\
DEEP-ITA-F   & 27.63 &26.79 & 27.20 & 0.86 & \textbf{0.75} &\textbf{0.81} \\
DEEP-ITA-S & \textbf{28.14} & {\textbf{27.38}}& {\textbf {27.76}} & {\textbf{0.86}}&  {{0.75}} & {0.81}     \\
\bottomrule
\end{tabular}
\end{sc}
\end{small}
\end{center}
\vskip -0.1in
\end{table*}
We compare Deep-ITA-F/S with other widely used algorithms on FPR, namely HIO \cite{fienup1982phase}, Oversampling Smoothness (OSS) \cite{rodriguez2013oversampling}, DnCNN-ADMM \cite{ venkatakrishnan2013plug, heide2016proximal, chan2016plug} and prDeep \cite{metzler2018prdeep}. We did not include any post-reconstruction procedure to clean the results as in \cite{icsil2019deep}, which is not tested here since the algorithm performs worse than prDeep unless an additional DNN specifically trained to enhance the quality is used. 

In principle, any denoiser can be adopted in RED. Here, we choose DnCNN \cite{zhang2017beyond} based on its competitive denoising performance and its flexibility on the input signal. DnCNN is stacked by Convolutional and Batch Normalization layers with Rectified Linear Unit (ReLU) activation functions. With padding of 1 for $3\times 3$ convolutional kernel size, the output dimension remains the same as that of the input. DnCNN models are trained on patches of natural images from  with mean-squared-error as the loss function using Adam as the optimizer \cite{kingma2014adam}.

The test images used in the simulations, shown in Figure.\ref{fig:testimages}, consist of 6 commonly used ``natural'' images and 6 ``unnatural'' ones. The images are resized to $128\times 128$ and their Fourier intensity are oversampled uniformly by a factor of $2$ in each dimension,  yielding measurements of size $256\times 256$. The signals used as ground truth are real-valued and have dynamic range of $[0, 255]$.

For simulation, shot noise is assumed to dominate the noise in the measurement. While this noise follows a Poisson distribution, it is commonly approximated as a Gaussian \cite{metzler2018prdeep, icsil2019deep}. The noisy measurement $y$ on the oversampled Fourier amplitude $q = \hat{x}^{(2)}$ thus has the distribution
\begin{equation}
    {y}^2 = |q|^2 + w\quad w\sim \mathcal{N}(0, \text{diag}(\alpha^2|q|^2))
\end{equation}
It is worth noting that the (effective) SNR in the measurements scales roughly with $y/\alpha$, which is affected by $\alpha$ and any scaling in $|q|$. We define two metrics to characterize the SNR: $\text{MSNR}_1=10\log_{10}(\rVert |q|^2\rVert_2/\rVert y^2 - |q|^2\rVert_2)$ \cite{icsil2019deep} and $\text{MSNR}_2=20\log_{10}(\rVert |q|\rVert_2/\rVert y - |q|\rVert_2)$ \cite{luke2004relaxed}. 

Results from two experimental setups are reported here. In the first, we test the convergence of the competing phase retrieval algorithms with random initialization. All algorithms are initialized with the same random point and run for the same total number of 1200 iterations. In the second, we follow the initializing strategy used in \cite{metzler2018prdeep, icsil2019deep}: first, make 50 runs of randomly initialized HIO (giving $\hat{x}_i$ for $i=1,\cdots, 50$), each with 50 iterations; next, pass the one with the lowest residual $\hat{x} = \text{argmin}_i f(\hat{x}_i)$ to initialize another HIO run of 1000 iterations. The output is then used as initialization for other algorithms. For both experiments, the whole procedure is repeated three times and the one most matched with the measurement is selected as the final output for each algorithm.

The parameters in the algorithms were as follows: for HIO and OSS, $\beta=0.9$. The regularization parameter $\lambda$ is found best set as $0.01\bar{\sigma}^2$ for DnCNN-ADMM, $0.025 \bar{\sigma}^2$ for both Deep-ITA-S/F, and $0.05 \bar{\sigma}^2$ for prDeep, where $\bar{\sigma}$ is the standard deviation of noise in the Fourier amplitude (or set to $0.1$ if no noise is added). Similar to the practice in \cite{metzler2018prdeep}, prDeep and Deep-ITAs sequentially use DnCNN models that are trained with noise standard deviations of 60, 40, 20 and 10, each with 300 iterations for a total of 1200 iterations. The penalty parameter $\rho$ used in Deep-ITAs is set to $\frac{1}{2}\lambda$. We notice that reducing $\lambda$ and $\rho$ when using the DnCNNs for high noise levels can increase the stability of our methods. We use the nonnegativity of the real part as the additional constraint $\mathcal{C}$ in the regularizer for prDeep and Deep-ITAs, which has the  element-wise projection
\begin{equation}
    \Pi_{Re_+}(x)_i = 
    \begin{cases}
    x_i &\text{if }\Re(x_i)\geq 0\\
    i\Im(x_i) &\text{otherwise}
    \end{cases}
\end{equation}
\begin{figure}[t]
    \centering
    \includegraphics[width=\linewidth]{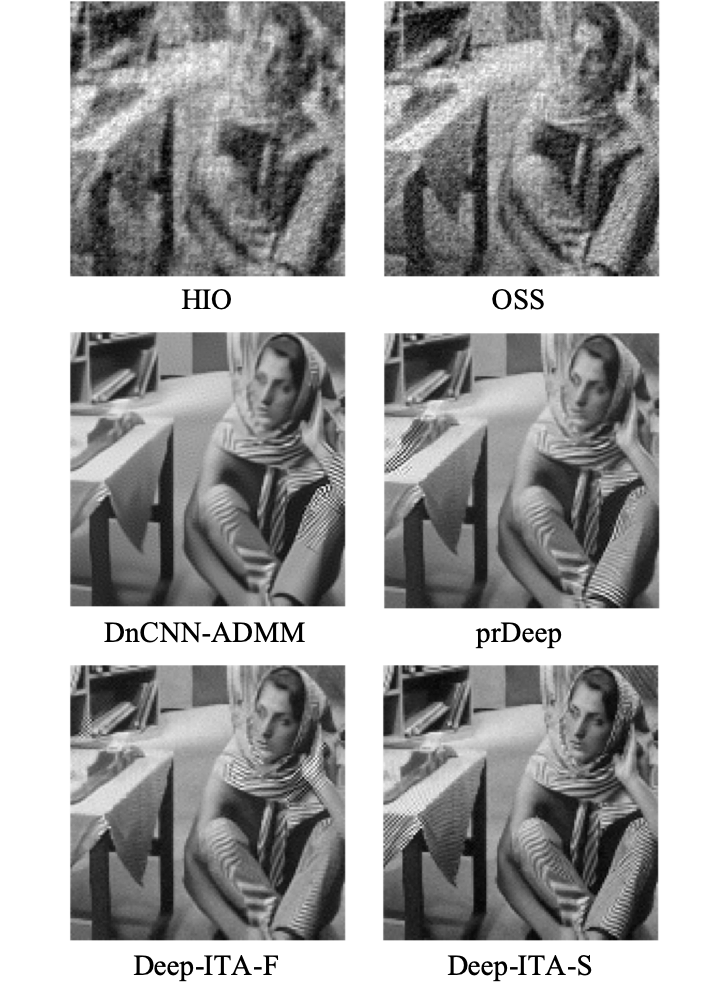}
    \caption{Reconstructions initialized from HIO with $\alpha = 12$. Deep-ITA-S has the best reconstruction results.}
    \label{fig:HIO_alpha12}
\end{figure}

For quantitative evaluation of the reconstructions, we characterize the output by its Peak Signal-to-Noise Ratio (PSNR) compared to the ground-truth as well as the Structure Similarity (SSIM) Index \cite{wang2004image}. The PSNR computed for each reconstruction is capped at $80$dB, in case an outlier has a high value and adversely affects the estimation of mean reconstruction quality (which could happen, e.g., in the noise-free case $\alpha=0$.)

\subsection{Random initialization}
Results of the experiments with random initialization are shown in Figure \ref{fig:rand_alpha4} and Table \ref{table_initializedrandom}. Our methods outperform every other PR algorithms by large margins, in both PSNR and SSIM. Significantly, this includes HIO even when noise is absent (Figure \ref{fig:rand_alpha0_stagnation}). (This is probably due to stagnation in HIO, which is hard to overcome in a limited number of iterations \cite{fienup1986phase}.) prDeep has issues with random initialization, which is not surprising considering its connection with Error Reduction, which has been shown to have slow convergence in practice \cite{fienup1982phase}. On the contrary, DnCNN-ADMM and Deep-ITAs have the ability to work with random initial points, since all of them use ADMM as a solver. Our methods are more effective, as we integrate the denoiser in the update via RED, rather than apply it in a Plug-and-Play manner.

\subsection{Initialization by HIO}
Table \ref{table_initializedHIO} shows the performance of test algorithms with different level of noise in Fourier intensity when initialized with HIO. Deep-ITAs exhibit higher robustness to noise for every level of noise added. Figure \ref{fig:HIO_alpha12} shows a visual comparison between PR algorithms for $\alpha = 12$, where Deep-ITA-S provides the best reconstruction. For the other methods, artifacts appear in the reconstructions and many details are lost.

%% file: conclusion.tex
\section{Conclusion}
Phase  retrieval  is  part  of  a  more  general  class  of  algorithms that has (to date) resisted full, end-to-end solutions
from machine learning.  While an admirable goal, such approaches often apply machine learning in situations where it is ill-suited.
It also neglects traditional algorithms and their corresponding strengths, viz. convergence benefits.

The approach advocated here is to build algorithms in the
fashion of traditional methods but with added priors utilizing
deep neural networks.   In the problem of Fourier phase
retrieval,  we added the object-space regularizer of image
statistics and improved noise robustness.
More generally, the results pave the way for hybrid methods
that integrate machine-learned constraints in conventional
algorithms.